%% file: main.tex
\documentclass[traditabstract]{aa}
\usepackage{graphicx}
\usepackage[intlimits,sumlimits]{amsmath}
\usepackage{times,epsfig} 
\usepackage{natbib}
\usepackage{amssymb}
\bibpunct{(}{)}{;}{a}{}{,} 
\usepackage{amsmath}
\usepackage[english]{babel}

\newcommand{\beqa}{\begin{eqnarray}} 

\newcommand{\eeqa}{\end{eqnarray}}

\newcommand{\bsub}{\begin{subequations}}
\newcommand{\esub}{\end{subequations}}
\newcommand{\beal}{\begin{align}}
\newcommand{\ealn}{\end{align}}
\newcommand{\lm}{$L_{max}$~}
\newcommand{\Mni}{$\rm M_{^{56}Ni}$~}

\newcommand{\mc}{$\rm M_{Ch}$~}
\newcommand{\Nif}{$
\rm ^{56}Ni$ } 
\newcommand{\Cif}{$\rm ^{56}Co$ }

\newcommand{\s}{M$_{\sun}$~}
\begin{document}
\title{A reddening-free method to estimate the \Nif mass of Type Ia supernovae
}

\titlerunning{Nickel mass in SNe~Ia}
\authorrunning{S. Dhawan et al.}
\author{\textbf{Suhail Dhawan\inst{1,2,3}
 	\and B. Leibundgut\inst{1,2}
    \and J. Spyromilio\inst{1}
    \and S. Blondin\inst{4}}}

\institute{European Southern Observatory, Karl-Schwarzschild-Strasse 2, D-85748 Garching bei M\"unchen, Germany \\
\email{sdhawan@eso.org}
\and  Excellence Cluster Universe, Technische Universit\"at M\"unchen,
Boltzmannstrasse 2, D-85748, Garching, Germany
\and Physik Department, Technische Universit\"at M\"unchen, James-Franck-Strasse 1, D-85748 Garching bei M\"unchen
\and Aix Marseille Universit\'e, CNRS, LAM (Laboratoire d'Astrophysique de Marseille) UMR 7326, 13388 Marseille, France
} 

\date{Received; accepted }

\offprints{S. Dhawan}

\abstract{The increase in the number of Type Ia supernovae (SNe\,Ia) has demonstrated that the population shows larger diversity than has been assumed in the past. The reasons (e.g. parent population, explosion mechanism) for this diversity remain largely unknown. We have investigated a sample of SNe\,Ia near-infrared light curves and have  correlated the phase of the second maximum with the bolometric peak luminosity. The peak bolometric luminosity is related to the time of the second maximum (relative to the {\it B} light curve maximum) as follows : $L_{max}(10^{43} erg s^{-1}) = (0.039 \pm 0.004) \times t_2(J)(days) + (0.013 \pm 0.106)$.

\Nif masses can be derived from the peak luminosity based on Arnett's rule, which states that the luminosity  at maximum is equal to instantaneous energy generated by the nickel decay. We check this assumption against recent radiative-transfer calculations of Chandrasekhar-mass delayed detonation models and find this assumption is valid to 
within 10\%  in recent radiative-transfer calculations of Chandrasekhar-mass delayed detonation models. 

The $L_{max}$ vs. $t_2$ relation is applied to a sample of 40 additional SNe\,Ia with significant reddening ($E(B-V) >$ 0.1 mag) and a reddening-free bolometric luminosity function of SNe~Ia is established. The method is tested with the \Nif mass measurement from the direct observation of $\gamma-$rays in the heavily absorbed SN~2014J and found to be fully consistent. 

Super-Chandrasekhar-mass explosions, in particular SN\,2007if, do not follow the relations between peak luminosity and second IR maximum. This may point to an additional energy source contributing at maximum light.

The luminosity function of SNe\,Ia is constructed and is shown to be asymmetric with a tail of low-luminosity objects and a rather sharp high-luminosity cutoff, although it might be influenced by selection effects.}

\maketitle
\section{Introduction}
\label{sec-intro}

Type Ia supernovae (SNe\,Ia) exhibit diverse observable properties. In addition to the spectral and colour differences, the peak luminosity of SNe\,Ia  range over several factors \citep[e.g.][]{Suntzeff1996, Suntzeff2003, Li2011}. The amount of $^{56}$Ni, derived from the bolometric luminosity, \citep{Contardo2000} and the total ejecta mass \citep{Stritzinger2006a, Scalzo2014} also show a wide dispersion. The $M_{^{56}Ni}$ distribution provides insight into the possible progenitor channels and explosion mechanisms for SNe\,Ia \citep[see][]{Hillebrandt2000, Livio2000, Truran2012}.

The uncertainty in the reddening correction directly impacts the ability to derive accurate bolometric luminosities and \Nif masses derived from the peak luminosity. 
The total to selective absorption ($R_V$) appears systematically and significantly lower in the SN hosts than the canonical Milky Way $R_V$ value of 3.1. \citet{Nobili2008} use a large sample of nearby SNe\,Ia to derive an average $R_V$ which is significantly lower than 3.1. Taking into account spectroscopic features that correlate with luminosity, \citet{Chotard2011} found an $R_V$ of $2.8 \pm 0.3$ which is consistent with the Milky Way value. However,  objects with high extinction are seen to have an unusually low $R_V$ \citep{Phillips2013, Patat2015}. Due to interstellar dust a correction for reddening in the Milky Way and the host galaxy needs to be applied. Our goal is to establish a relation between the bolometric peak luminosity and the \Nif mass independent of reddening.

The Near Infrared (NIR) light curve morphology of SNe\,Ia is markedly different from that in the optical. In particular, the light curves start to rebrighten about 2 weeks after the first maximum, resulting in a second peak. 
Recent studies \citep[e.g.][]{Biscardi2012,Dhawan2015} found that more luminous SNe\,Ia reach the second maximum in NIR filters at a later phase. This was predicted by \citet{Kasen2006} who also indicated that the phase of the second maximum (designated in the following as $t_2$  and measured relative to the $B$-band light curve maximum) should be a function of the \Nif mass in the explosion. We expect that the phase of the second NIR maximum can be used to determine bolometric peak luminosity $L_{max}$ and the amount of \Nif  synthesized in the explosion. 

In the following, we investigate the link between the peak bolometric luminosity ($L_{max}$) and the phase of the second maximum in the NIR light curves ($t_2$). We use a sample of nearby SNe\,Ia with low host-galaxy extinction (described in Section~\ref{sec-data}) to determine $L_{max}$ and then employ different methods to derive $M_{^{56}Ni}$ (Section~\ref{sec-res}). This relation can then be used to derive $L_{max}$ and $M_{^{56}Ni}$ for all SNe\,Ia with a measured $t_2$, since the timing parameter is free of reddening corrections and allows us to include heavily reddened objects. We check our derivation against independent measurements of $M_{^{56}Ni}$ with the nearby SN~2014J in M82 and SN~2006X (Section~\ref{sec-test}). With the reddening independent method we can establish the luminosity function of SNe~Ia at maximum and also derive the distribution of nickel masses among SN~Ia explosions (Section~\ref{sec-lfunc}). We finish by discussing the implications of this determination of the $M_{^{56}Ni}$ distribution in the conclusions (Section~\ref{sec-dnc}).

\begin{figure}
\centering
\includegraphics[width=.50\textwidth, height=0.6\textheight]{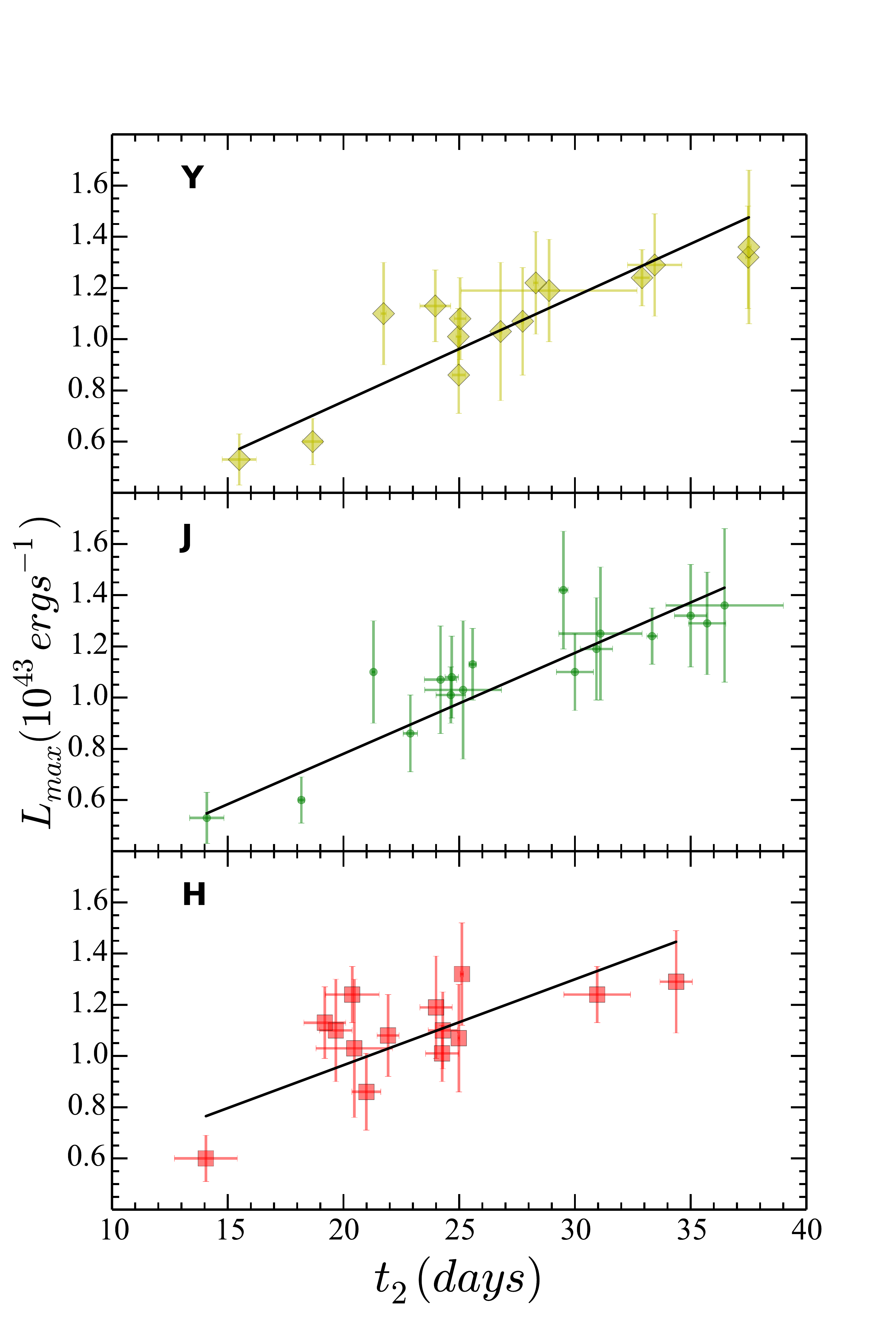}
\caption{The bolometric maximum luminosity $L_{max}$ is plotted against the phase of the second maximum $t_2$ in $YJH$ filter light curves. A strong correlation is observed in  $Y$ and $J$, whereas a weaker correlation is seen in the $H$ band.  Best fit lines are overplotted in black. The fit includes errors on both axes.} 
\label{fig:nit2}
\end{figure}

\section{Data} 
\label{sec-data}

Our SN\,Ia sample is constrained to objects, which have NIR observations at late times ($t>50$ days after B maximum) with well-sampled optical and NIR light curves to construct a (pseudo-)bolometric light curve and measure $t_2$. The main source of near-infrared photometry of SNe\,Ia is the Carnegie Supernova Project \citep[CSP;][]{ Contreras2010,Burns2011,Stritzinger2011,Phillips2012,Burns2014}. 
We add to this sample objects from the literature. We only included SNe\,Ia with observations near maximum from $U$ to $H$ filters. The full description of the selected SNe\,Ia can be found in \citet{Dhawan2015}.

The sample of low-reddening SNe\,Ia is defined to circumvent the uncertainties of host galaxy extinction. The 18 objects are presented in Table~\ref{tab:lr}. We use $E(B-V)_{host}$ values from the literature.   Only objects with $E(B-V)_{host}<0.1$ mag were included. Since we consider only objects, which display the second maximum in their NIR light curves this implies that most low-luminosity SNe\,Ia, especially SN~1991bg-like objects were excluded.

At maximum light the UV-optical-IR integrated luminosity represents $>$90\% of the true bolometric luminosity \citep{Blondin2015}.
We constructed $UBVRIJH$ bolometric light curves for objects with sufficient photometry near maximum light in the optical and the NIR. The $K$ filter data was excluded since only few SN~Ia have well-sampled $K$ light curves. We calculated the fraction of the flux emitted in $K$ for a few well-observed SNe~Ia with sufficient data and determined it to be around $1-3 \%$ of the UVOIR luminosity at maximum. The exclusion of the $K$-flux results in only a minor uncertainty in the final UVOIR luminosity. 

Prior to the  derivation of a bolometric flux for the low extinction sample (see Table~\ref{tab:mni}) we apply a correction for the measured extinction following \citet{Cardelli1989}. The assumed distances and their references can be found in Table~\ref{tab:lr}.

\input{tab/sn_samp_lr.tex}


\section{Results}
\label{sec-res}
\input{res.tex}

\section{Discussion and Conclusion}
\label{sec-dnc}
\input{dis.tex}


\begin{acknowledgements}
This research was supported by the DFG Cluster of Excellence ʻOrigin and
Structure of the Universe'.
B.L. acknowledges support for this work by the Deutsche
Forschungsgemeinschaft through TRR33, The Dark Universe. We all are grateful to the ESO Visitor Programme to support the visit of S. B. to Garching.
We would like to thank Rahman Amanullah for providing published photometry of SN2014J in the near infrared. 

\end{acknowledgements}

\input{ref.tex}
\end{document}

%% file: tab/sn_samp_lr.tex
\begin{table*}\caption{The sample of SNe~Ia with low reddening $E(B-V)_{host}<0.1$. The references for the SNe~Ia are presented along with the extinction values and the distances employed for calculation of the (pseudo-) bolometric luminosity at maximum.}

\begin{center}
\begin{tabular}{llccccccc}
\hline
SN  & \multicolumn{1}{c}{$\mu$} & $E(B-V)_{host} $ & $E(B-V)_{MW}$ & Reference & $t_2(Y)$ & $t_2(J)$ & $t_2(H)$ & $L_{max} $\\
 & & & & & \multicolumn{1}{c}{(d)} & \multicolumn{1}{c}{(d)} & \multicolumn{1}{c}{(d)} & \multicolumn{1}{c}{($10^{43}$ erg/s)}\\
\hline

SN2002dj & 32.93 $\pm$  0.30 & $ 0.096 \pm 0.030$ & 0.010 $\pm$ 0.003 & P08 & \ldots	& $31.1 \pm 1.8$ 	&23.0	 $\pm$ 	0.9	 & 1.25 $\pm$ 0.26\\
SN2002fk & 32.59 $\pm$ 0.15 & $0.009 \pm 0.044$ & 0.035 $\pm$ 0.003 & C14 & \ldots	& $29.5 \pm 0.2$  	&25.8	 $\pm$ 	0.3	& 1.42 $\pm$ 0.23 \\
SN2005M  & 35.01 $\pm$ 0.09 & $0.060 \pm 0.021$ & 0.027 $\pm$ 0.002 & B14 & $28.9 \pm 3.8$ 	& $30.9 \pm 0.7$ 	&\ldots	 & 1.19 $\pm$ 0.20 \\
SN2005am & 32.85 $\pm$ 0.20 & $0.053 \pm 0.017$ & 0.043 $\pm$ 0.002 & B14 & $21.7 \pm 0.1$ 	& $21.3 \pm 0.7$	&19.7	 $\pm$ 	0.7	& 1.10 $\pm$ 0.20 \\
SN2005el & 34.04 $\pm$ 0.14	& $0.015 \pm 0.012$ & 0.098 $\pm$ 0.001 & B14 & $25.0 \pm 0.1$	& $24.6 \pm 0.6$ 	&24.3	 $\pm$ 	0.7  & 1.01	$\pm$ 0.11	 \\
SN2005eq & 35.46 $\pm$ 0.07 & $0.044 \pm 0.024$ & 0.063 $\pm$ 0.003 & B14  & $37.5 \pm 0.1$	& $35.0 \pm 0.7$ 	&25.1	 $\pm$ 	0.1	& 1.32 $\pm$ 0.20  \\
SN2005hc & 36.50 $\pm$ 0.05 & $0.049 \pm 0.019$ & 0.028 $\pm$ 0.001 & B14	& $37.5 \pm 0.1$	& $36.5 \pm 2.5$ 	&\ldots	& 1.36 $\pm$ 0.30 \\
SN2005iq & 35.80 $\pm$ 0.15 & $0.040 \pm 0.015$ & 0.019 $\pm$ 0.001 & B14	& $27.7 \pm 0.1$ 	& $24.2 \pm 0.7$ 	&25.0	 $\pm$ 	0.1	& 1.07 $\pm$ 0.21 \\
SN2005ki & 34.73 $\pm$ 0.10 & $0.016 \pm 0.013$ & 0.027 $\pm$ 0.001 & B14	& $26.8 \pm 0.1$ 	& $25.2 \pm 1.7$ 	&20.5	 $\pm$ 	1.7	& 1.03 $\pm$ 0.27  \\
SN2006bh & 33.28 $\pm$ 0.20 & $0.037 \pm 0.013$ & 0.023 $\pm$ 0.001 & B14	& $25.0 \pm 0.3$ 	& $22.9 \pm 0.3$ 	&21.0	 $\pm$ 	0.6	& 0.86 $\pm$ 0.15  \\
SN2007bd & 35.73 $\pm$ 0.07 & $0.058 \pm 0.022$ & 0.029 $\pm$ 0.001 & B14	& $28.3 \pm 0.1$	& \ldots		&\ldots	 & 1.22 $\pm$ 0.13 \\
SN2007on & 31.45 $\pm$ 0.08 & $<0.007$ 	& 0.010 $\pm$ 0.001 & B14	& $18.7 \pm 0.4$	& $18.2 \pm 0.1$ 	&14.1	 $\pm$ 	1.4	  & 0.60 $\pm$ 0.09 \\
SN2008R  & 33.73 $\pm$ 0.16 & $0.009 \pm 0.013$ & 0.062 $\pm$ 0.001 & B14	& $15.5 \pm 0.7$	& $14.1 \pm 0.7$ 	&\ldots	& 0.53 $\pm$ 0.10  \\
SN2008bc & 34.16 $\pm$ 0.13 & $<0.019$ 	& 0.225 $\pm$ 0.004 & B14	& $32.9 \pm 0.3$	& $33.3 \pm 0.2$ 	&31.0	 $\pm$ 	1.4	& 1.24 $\pm$ 0.19 \\
SN2008gp & 35.79 $\pm$ 0.06 & $0.098 \pm 0.022 $ & 0.104 $\pm$ 0.005 & B14	& $33.5 \pm 1.2$	& $35.7 \pm 0.8$ 	&34.4	 $\pm$ 	0.7	& 1.29 $\pm$ 0.14  \\
SN2008hv & 33.84 $\pm$ 0.15 & $0.074 \pm 0.023 $ & 0.028 $\pm$ 0.001 & B14	& $25.0 \pm 0.3$ 	& $24.7 \pm 0.3$ 	&21.9	 $\pm$ 	0.5	& 1.08 $\pm$ 0.16 \\
SN2008ia & 34.96 $\pm$ 0.09 & $0.066 \pm 0.016$ & 0.195 $\pm$ 0.005 & B14	& $24.0 \pm 0.7$	& $25.6 \pm 0.2$ 	&19.2	 $\pm$ 	0.9	& 1.13 $\pm$ 0.14 \\
SN2011fe & 28.91 $\pm$ 0.20 & $0.014 \pm 0.010$ & 0.021 $\pm$ 0.001 & P13 & \ldots	& $30.0 \pm 0.8$ 	&24.3	 $\pm$ 	0.6	& 1.10 $\pm$ 0.15 \\
\hline
\multicolumn{9}{l}{{\bf E(B-V) references:}
P08: \citet{Pignata2008}; C14: \citet{Cartier2013} B14:\citet{Burns2014};P13: \citet{Patat2013}}\\

\end{tabular}
\label{tab:lr}
\end{center}
\end{table*}

%% file: res.tex
Based on our previous work \citep{Dhawan2015}, where we found strong correlations between various derived parameters of SNe\,Ia with $t_2$ in the $Y$ and $J$ filters, we argued that the bolometric maximum luminosity should also correlate with $t_2$. The sample of low-reddening SNe\,Ia described in Section~{\ref{sec-data} is used to establish the relation between $t_2$ and $L_{max}$. The $t_2$ parameter has the advantage that is is essentially independent of reddening and distance (relative to the calibration sample). With such a relation we will be in a position to derive the luminosity function of SNe\,Ia at maximum. 


\subsection{Correlation between $L_{max}$ and $t_2$}

Figure ~\ref{fig:nit2} displays a strong correlation between $t_2$ for the $Y$ and $J$ filter light curves and the bolometric (UVOIR) luminosity $L_{max}$ (determined by fitting a spline interpolation to the UVOIR light curve) with Pearson coefficients $r=0.88$ and $r=0.86$, respectively for the low-reddening sample. A much weaker trend is observed in the $H$ filter light curve with $r$$\approx$0.60. 

The slope of the $L_{max}$ vs. $t_2$ relation appears to flatten for objects with $t_2 \gtrsim$ 27\,d. This is most prominent in the $Y$ band. However, we would require a larger sample to confirm this trend amongst the most luminous objects.

We fit a simple linear relation to the data
\begin{equation}
\label{eq:lmt2}
L_{max}=a_i \cdot t_{2,i} + b_i
\end{equation}
which leads to the entries in Table ~\ref{tab:coeff} (for $YJH$ filters). The corresponding fits are shown in Figure~\ref{fig:nit2}. We compare our observed values to the \lm and $t_2$ from the DDC models of \citet{Blondin2013}. Model spectra and light curves published in \citet{Blondin2013} based on delayed detonation explosions show similar correlations to those described herein.

In the interest of a clean low extinction sample, we have removed 7 objects with $E(B-V)_{host}<0.1$ but total $E(B-V) \geq 0.1$. Interestingly, several of the excluded objects are amongst the most luminous SNe~Ia in the sample. 
Even after the removal of these 7 objects we do not derive a significant correlation for the $H$ band light curves from our sample. It will have to be seen, whether future data will reveal a correlation or whether the $H$ light curves are not as sensitive to the nickel mass as the other NIR filters. The relations are identical for the full and restricted sample within the uncertainties listed in Table~\ref{tab:coeff}. We combine the relations from the two bands for extrapolating the values of $L_{max}$ in the following analysis. We assume the $Y$ band estimate to be equivalent to the value in the $J$ band and calculate the slope and intercept with the photometry of both filters, which leads to improved statistics.

\input{tab/coeff}

\subsection{Deriving $M_{^{56}Ni}$ from $L_{max}$}

\label{ssec-derni}
\input{mni_tab.tex}

We present three different methods to derive $M_{^{56}Ni}$ from $L_{max}$, namely using Arnett's rule with an individual rise time for each SN~Ia, using Arnett's rule with an assumed constant rise time applied to all SNe~Ia and by calculating $L_{max}$ from delayed detonation models with different \Mni yields \citep{Blondin2013}.
Arnett's rule states that at maximum light the bolometric luminosity equals the instantaneous rate of energy input from the radioactive decays. Any deviations from this assumption are encapsulated in a parameter $\alpha$ below. It is quite possible that $\alpha$ depends on the explosion mechanism and shows some variation between explosions \citep{Branch1992, Khokhlov1993}. These early papers found rather large ranges with $0.75 < \alpha < 1.4$ depending on the exact explosion model and the amount of assumed mixing \cite{Branch1992, Khokhlov1993}. More recently \citet{Blondin2013} found a range of $\alpha$ within 10\% of 1 for delayed detonation models. These models are not applicable for low-luminosity SNe\,Ia. Should $\alpha$ systematically depend on explosion characteristics then the derived nickel masses may suffer from a systematic drift not captured in our treatment. These uncertainties must be taken into account for the interpretation of the derived $^{56}$Ni mass.

\subsubsection{Arnett's rule with individual rise times}

Arnett's rule states that the luminosity of the SN at peak is given by the instantaneous rate of energy deposition from radioactive decays inside the expanding ejecta \citep{Arnett1982, Arnett1985}. 

This is summarized as \citep{Stritzinger2006a}: 
\begin{equation}
\label{eq:lm-eni}
L_{max}(t_R)=\alpha E_{^{56}Ni}(t_R),
\end{equation}
where $E_{^{56}Ni}$ is the rate of energy input from $^{56}$Ni and $^{56}$Co decays at maximum, $t_R$ is the rise time to bolometric maximum and $\alpha$ accounts for deviations from Arnett's Rule. The energy output from 1 \s of \Nif is ($\alpha=1$):

\begin{equation}
\label{eq:eni}
\epsilon_{Ni} (t_R, 1~M_{\odot})= \\(6.45  \cdot 10^{43} e^{-t_R/8.8} + 1.45  \cdot  10^{43} e^{-t_R/111.3}) {\rm erg s^{-1}}
\end{equation}

We use the relation for estimates using different rise times in the $B$ filter for each SN following,  
\begin{equation}
t_{R, B}=17.5 - 5 \cdot (\Delta m_{15} - 1.1)
\end{equation}
from \citet{Scalzo2014} which covers the $t_{R,B}$--$\Delta m_{15}$ parameter space of \citet{G11}. Like \citet{Scalzo2014} we apply a conservative uncertainty estimate of $\pm 2$~days.
The bolometric maximum occurs on average 1~day before $B_{max}$ \citep{Scalzo2014}.

\subsubsection{Arnett's rule with a fixed rise time}

Originally $M_{^{56}Ni}$ was determined from $L_{max}$ for a fixed rise time of 19 days for all SNe~Ia \citep{Stritzinger2006a}. Similar to these analyses we propagate an uncertainty of $\pm$3~days to account for the diversity in the rise times. 

The peak luminosity then becomes \citep{Stritzinger2006a} 
\begin{equation}
\label{eq:arn}
L_{max}=(2.0 \pm 0.3) \cdot 10^{43} (M_{^{56}Ni}/M_{\odot}) {\rm erg s^{-1}}.
\end{equation}

As described above, we assumed $\alpha=1$ \citep[see][]{Stritzinger2006a, Mazzali2007}, which is the analytical approximation of \citet{Arnett1982}. For the DDC models of \citet{Blondin2013} $\alpha$ is within 10 $\%$ of 1 for all but the least luminous model.


\subsubsection{Interpolating using delayed detonation models}
We interpolate the relation between \lm (in a given filter set, u $\rightarrow$ H in this case) and \Mni found from a grid of Chandrasekhar mass delayed detonation models of \citet{Blondin2013} to derive a \Nif mass from the  observed peak luminosity for the sample presented in Table \ref{tab:lr}.
The resulting \Nif mass estimates are presented in the bottom panel of Figure \ref{fig:nicomp}.
For all but the least luminous of these models, $\alpha$ is within 10 $\%$ of 1 \citep{Blondin2013}. 

\begin{figure}
\includegraphics[width=.5\textwidth, trim= 0 0 0 30]{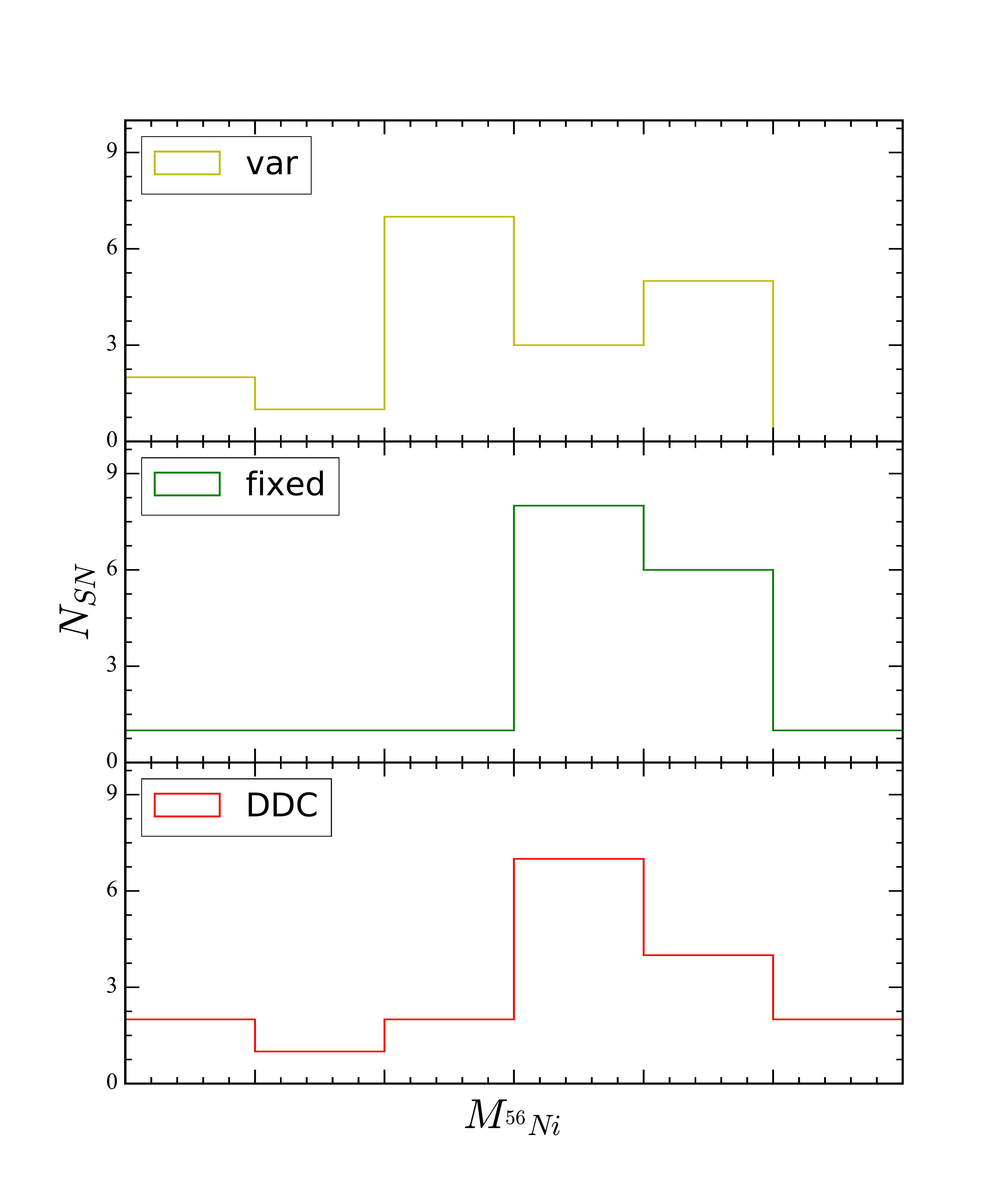}
\caption{The histograms show the different methods to estimate the $M_{^{56}Ni}$ from the $L_{max}$. The values found from Arnett's rule with fixed and individual rise times are plotted in the \emph{top} and \emph{middle} panels. The \emph{bottom} panel displays the values estimated by using the relation between \lm and \Mni found from the DDC models.}
\label{fig:nicomp}
\end{figure}

\subsection{Comparison of different methods}
In Figure~\ref{fig:nicomp}, we plot the distributions of the $M_{^{56}Ni}$, inferred for the low-reddening sample, from the different methods.

Similar to previous studies we find that there is a large distribution in the $M_{^{56}Ni}$ values for the sample in Table \ref{tab:lr}. We note a over a factor of 2 difference between the lowest and highest $M_{^{56}Ni}$ values (a factor of $\sim$ 3 for the variable rise time formalism). Unlike previous studies, this sample doesn't include faint, 91bg-like objects, since their NIR light curves do not display a second maximum. These objects are found to have a much lower $M_{^{56}Ni}$$\sim$0.1 $M_{\odot}$ \citep{Stritzinger2006a, Scalzo2014} from their peak luminosities. There is clearly a majority of objects with nickel masses between 0.4 and 0.7~$M_\odot$ with extensions to higher and lower masses. A further difference can be seen in Fig.~\ref{fig:nicomp} where the mass distribution between the case of individual and the fixed rise times is slightly different due to the fact that observed rise times often are shorter than the assumed 19 days.

The individual errors clearly dominate over the differences between the methods and the results are not influenced by the chosen method. There appears a small systematic offset between the \Nif masses derived from DDC models and the ones with Arnett's rule and fixed rise time. The \Nif masses from the DDC models are about 0.05~$M_\odot$ smaller, however, well within the overall uncertainties, which are typically around 0.15~$M_\odot$ (Tab.~\ref{tab:comp_ni}). 

For SN2011fe \citet{Pereira2013} report a rise time of 16.58\,days. Using this rise time, we obtain an \Mni of 0.49 \s which is a 0.06 \s shift from the value using a 19\,d rise. This shift is smaller than the uncertainty on the \Nif mass. For the low-reddening sample we note that the average difference between the fixed rise and variable rise formalisms is 0.04~M$_\odot$. 

For the following analysis, we calculate the \Mni using $t_2$. By substituting the relation derived between $L_{max}$ and $t_2$ (equation \eqref{eq:lmt2} and equation \eqref{eq:eni}), we obtain
\begin{equation}
\label{eq:nit2}
\frac{M_{^{56}Ni}}{M_\odot} = \frac{a_i \cdot t_2(i) + b_i}{\alpha \cdot \epsilon_{Ni}(t_R, M_\odot)} .
\end{equation}

We mostly will use the fixed rise time formalism in the following analysis, although in special cases, we will also make use of the more accurately known rise time. 



\section{Test with well observed SNe~Ia}
\label{sec-test}
Dust in the host galaxy and the Milky Way makes the determination of the peak luminosity uncertain. Many nearby SNe~Ia have shown marked deviations in the host galaxy dust properties from those observed in the Milky Way mostly favouring a smaller $R_V$ value \citep{Goobar2008, Phillips2013}. The extinction corrections are notoriously uncertain and directly affect our ability to measure peak bolometric luminosities of SNe~Ia. Since $t_2$ is independent of reddening, we can use the  derived correlation to determine the peak luminosity and estimate the \Nif mass for heavily reddened SNe~Ia. 

We test this relation on SN~2014J, which has a direct $\gamma-$ray detection from the \Nif $\rightarrow$ \Cif decay chain \citep{Churazov2014, Diehl2015}. Using the best fit relation for the reddening-free sample, we obtain $M_{^{56}Ni} = 0.64 \pm 0.15 M_{\odot}$ for a $t_2 = 31.99 \pm 1.15$ days and a rise time of 19~days. Since the error on the rise time is taken as $\pm$3 days, we expect the error on $M_{^{56}Ni}$ to decrease with a less conservative error estimate on $t_R$. \citet{Goobar2014b} used Palomar Transient Factory (PTF) and Kilodegree Extremely Little Telescope (KELT) data to measure the rise time of SN~2014J. They find $t_R = 17.25$~days. We place a conservative error estimate of 1 day and evaluate the $M_{^{56}Ni} = 0.60 \pm 0.10 M_{\odot}$ which has a lower error bar than from the fixed rise time formalism.

The direct measurement of $M_{^{56}Ni}$ for SN~2014J through the $\gamma-$ray detection gives an independent and fairly secure estimate of the nickel mass. \citet{Churazov2014} derive $^{56}Ni = 0.62 \pm 0.13 M_{\odot}$. \citet{Diehl2015} find a slightly lower mass of $M_{^{56}Ni} = 0.56 \pm 0.10 M_\odot$.

\input{tab/14j_meth}

A detailed comparison of the derived \Nif masses is given in Table~\ref{tab:meth}. The difficulty of the extinction correction and the advantage of the method presented here are obvious. The uncertainty in the $\gamma-$ray determination is due to the weakness of the signal and leads to slightly different interpretations. The very good correspondence between the direct \Mni measurement and our relation of the second maximum in the NIR light curves is encouraging. 

As a second case, we determine the bolometric peak luminosity $L_{max}$ and the nickel mass $M_{^{56}Ni}$ based on $t_2$ to the heavily extinguished SN~2006X. \citet{Wang2008} derived a peak luminosity for this SN from multi-band photometry and a correction for dust absorption in the host galaxy. They determined a bolometric peak luminosity for SN\,2006X of $(1.02 \pm 0.1) \cdot 10^{43}$ erg s$^{-1}$, which compares well with our measurement of $(1.14 \pm 0.16) \cdot 10^{43}$ erg s$^{-1}$. \citet{Wang2008} determined $M_{^{56}Ni} = 0.50 \pm 0.05 M_{\odot}$, which should be compared to $M_{^{56}Ni} = 0.57 \pm 0.11$ $M_{\odot}$ found from $t_2$ using the fixed rise time formalism. The measured rise time for SN~2006X is $t_{R}(B) = 18.2 \pm 0.9$\,d, which leads to $M_{^{56}Ni}= 0.55 \pm 0.10$~M$_\odot$.

\input{red_tab}

Table~\ref{tab:red} presents several additional highly reddened SNe~Ia, which had a previous determination of the nickel mass. The $M_{^{56}Ni}$ for these objects were calculated in the same way as for SN~2014J and SN~2006X. 

From Table ~\ref{tab:red}, we can see that 1986G has a lower value of $M_{^{56}Ni}$ than the other heavily reddened objects. This is consistent with the observed optical decline rate and lower $B$ band luminosity \citep{Phillips1987}. Using nebular spectra, \citet{RL1992} calculate the \Mni\ for SN~1986G and find a value of $0.38 \pm 0.03$ \s. This is fully consistent with the estimate from $t_2$.  

\citet{Scalzo2014} give M$_{^{56}Ni}$ for SN~2005el and SN~2011fe. The comparison for SN~2011fe shows $M_{^{56}Ni} = 0.52 \pm 0.15$ \s from the NIR light curves, whereas \citet{Scalzo2014} find $M_{^{56}Ni} = 0.42 \pm 0.08$ \s. The difference is mostly in the adopted value of $\alpha$, 1.2 in \citet{Scalzo2014}  compared to 1 in this study.
Rescaling the value from \citet{Scalzo2014} to $\alpha$=1, we obtain \Mni = 0.50 $\pm$ 0.08 \s, which is fully consistent with our value. \citet{Pereira2013} report nickel masses for SN~2011fe for different values of $\alpha$. Their nickel mass for $\alpha$=1 is $M_{^{56}Ni} = 0.53 \pm 0.11$ \s, nearly identical to our determination. 
For SN~2005el, \citet{Scalzo2014} obtain an \Mni of 0.52 $\pm$ 0.12\,M$_\odot$. Scaled to an $\alpha=1$, this gives \Mni\ = 0.62. We find \Mni\ = 0.51 $\pm$ 0.11\,M$_\odot$, which is broadly consistent with the value found in \citet{Scalzo2014}.

From the comparisons in Table ~\ref{tab:red}, we conclude that there is good agreement between our values and those published in the literature. For SN~2001el we see that the error in the estimate from $t_2$ is substantially smaller than from the bolometric light curve.

One significant outlier is SN\,2007if. This was presented as a super-Chandrasekhar-mass explosion \citep{Scalzo2010} with a total luminosity of $3.2 \cdot 10^{43}$ erg s$^{-1}$. The reddening from the host galaxy is somewhat unclear. There is no indication of Na foreground absorption, while the colour evolution and the Lira law would indicate some reddening. Any reddening would only increase the luminosity and the derived nickel mass based on Arnett's rule. The \Mni estimate from $t_2$ for SN\,2007if is significantly lower than the mass estimate through the bolometric peak luminosity by \citet{Scalzo2010}. If we recalculate the \Mni from the bolometric light curve presuming no extinction from the host galaxy, we obtain \Mni = 1.6\,M$_\odot$. This is a factor of $\sim$ 2 larger than our estimate. We discuss this supernova in Section~\ref{sec-dnc}.


\section{The luminosity function of SNe~Ia at maximum}
\label{sec-lfunc}

We are now in a position to derive $L_{max}$ for all SNe~Ia with a reliably measured $t_2$ (as given in Tables~\ref{tab:lr} and \ref{tab:yj}) and establish the bolometric luminosity function of SNe~Ia at maximum light. For objects in the low-reddening sample, we use the \lm determined from $t_2$ and the best-fit linear relation (green histogram in Figure ~\ref{fig:hist}). 
Since the phase of the second maximum in the near infrared is independent from the reddening we can derive the reddening-free distribution of the luminosity function of SNe~Ia (Fig.~\ref{fig:hist}). We show here the histogram of 58 SNe~Ia as derived from the $Y$ and $J$ light curves. The luminosity scale is based on the calibration sample of low-reddening objects (Section~\ref{sec-res}).

The luminosity function of SNe~Ia is clearly not symmetric. The luminosity range spans slightly over a factor of 2. We find no obvious difference between the full sample and the low-reddening sample used to calibrate the relation between $t_2$ and $L_{max}$. If anything the calibration sample has a flatter distribution with most SNe around $0.9\cdot10^{43}$~erg s$^{-1}$, while the full sample includes more luminous objects. This could be an effect of the magnitude limit of the searches. The exact biases in our sample are difficult to define as it is not volume limited. 

\begin{figure}
\includegraphics[width=.5\textwidth, trim= 0 30 0 30]{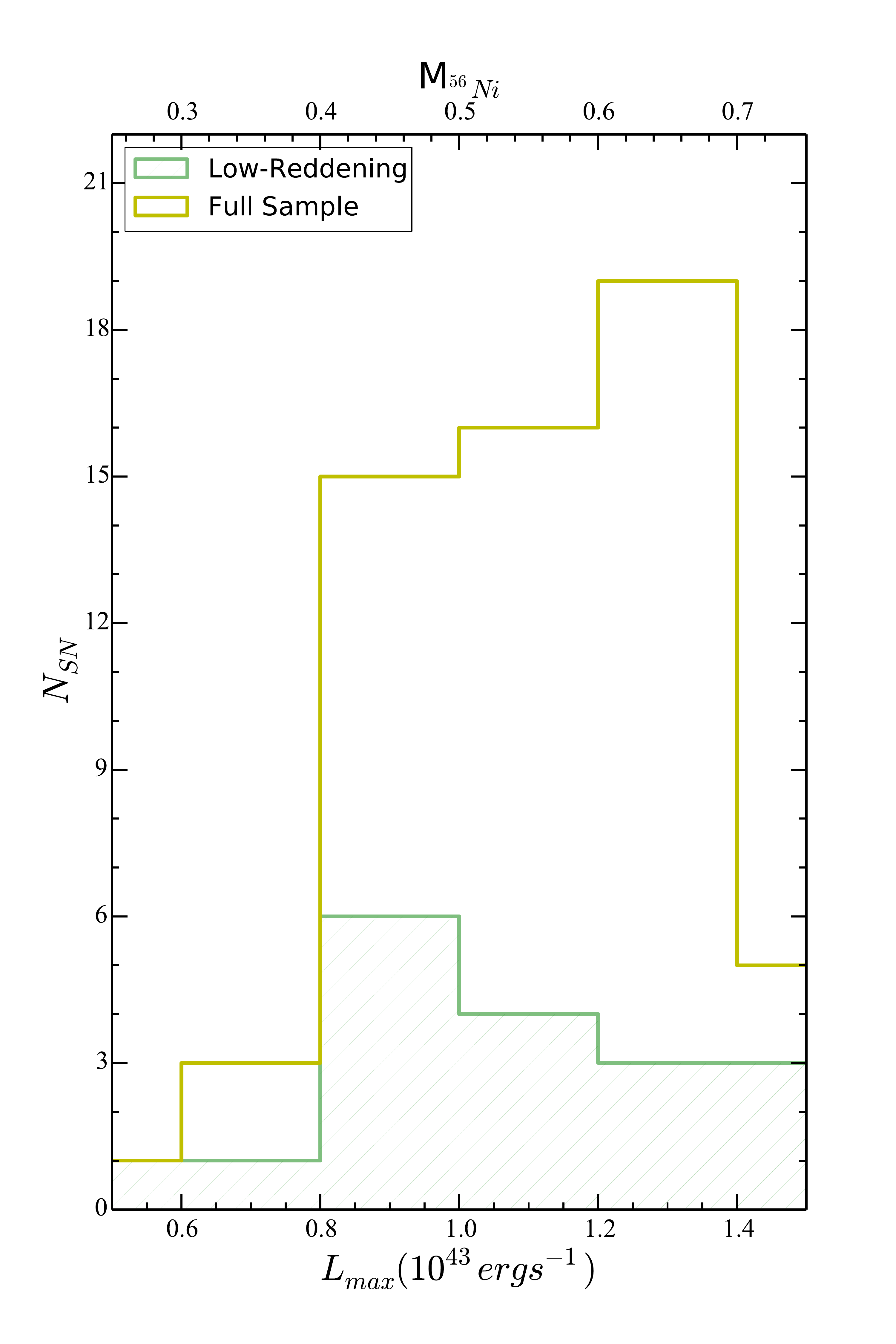}
\caption{Histogram distribution of \lm derived from the relations for the full sample of objects.  \emph{Green:} Estimates of \lm for the low-reddening, calibration sample. \emph{Yellow:} Estimates for all objects with a $t_2$ measurement in the full sample.
The axis labels on top correspond to the \Mni estimate. We use the combined fit to obtain the final values.}.
\label{fig:hist}
\end{figure}

The bolometric luminosity function can be compared to the $R$ filter luminosity function derived by \citet{Li2011} based on 74 SNe~Ia including the low-luminosity objects missing in our sample. The \citet{Li2011} magnitude-limited luminosity function (their Fig.~10) peaks at an absolute magnitude of $\approx -19$ with a few objects above $-19.5$ and a tail to fainter objects down to $-17$. This is also
reflected in our luminosity function (Fig.\ref{fig:hist}), where we observe a clear peak at $L_{max} = 1.3 \cdot 10^{43} {\rm erg s^{-1}}$ with some more luminous objects and a tail to fainter objects. The range is also comparable to the one found by \citet{Li2011}. 

In the next step we derive the distribution of \Mni for all SNe~Ia with sufficient infrared light curve data using equation~\ref{eq:nit2} and a fixed rise time and $\alpha=1$. Table~\ref{tab:yj} and Fig.~\ref{fig:hist} present the SN~Ia nickel mass function. 

\input{nitable6}

%% file: tab/coeff.tex
\begin{table}
\caption{Values of the coefficients for correlations between $L_{max}$ and $t_2$ in the individual filters}
\begin{center}
\begin{tabular}{lrr}
\\
\hline

Filter & \multicolumn{1}{c}{$a_i$} & \multicolumn{1}{c}{~~~~$b_i$}\\
\hline

Y    &	$0.041 \pm 0.005$  &	$-0.065 \pm 0.122$\\
J    & 	$0.039 \pm 0.004$  &	$ 0.013 \pm 0.106$\\
H    & 	$0.032 \pm 0.008$  &	$ 0.282 \pm 0.174$\\
\hline
\end{tabular}
\end{center}
\label{tab:coeff}
\end{table}

%% file: mni_tab.tex
\begin{table*}
\caption{\Mni\ measurements for low reddening SNe~Ia. The components of the error from $L_{max}$ and rise time are given along with the total error.}
\label{tab:comp_ni}
\begin{center}
\begin{tabular}{llcccccc}
\hline
SN  & $M_{Ni}-Arn$  & err ($L_{max}$)& err (rise time)& $M_{Ni}-Arn $ (fixed rise)  & err ($L_{max}$) & err (rise time) & $M_{Ni}-DDC $ \\
 & \multicolumn{1}{c}{(\s)} & \multicolumn{1}{c}{(\s)} & \multicolumn{1}{c}{(\s)} & \multicolumn{1}{c}{(\s)} & \multicolumn{1}{c}{(\s)} & \multicolumn{1}{c}{(\s)} & \multicolumn{1}{c}{(\s)} \\
\hline
SN2002dj  & 0.59 $\pm$ 0.16 & 0.12 & 0.10 & 0.63 $\pm$ 0.16 & 0.13 & 0.10 & 0.61 $\pm$ 0.13 \\
SN2002fk  & 0.68 $\pm$ 0.16 & 0.11 & 0.12 & 0.71 $\pm$ 0.17 & 0.12 & 0.12 & 0.76 $\pm$ 0.13 \\
SN2005M  & 0.59 $\pm$ 0.14 & 0.10 &  0.10 & 0.60 $\pm$ 0.14 & 0.10 & 0.10 & 0.59 $\pm$ 0.11 \\
SN2005am  & 0.47 $\pm$ 0.13 & 0.08 &  0.10 & 0.55 $\pm$ 0.14 & 0.10 & 0.10 & 0.52 $\pm$ 0.11 \\
SN2005el & 0.45 $\pm$ 0.10 & 0.05 &  0.09 & 0.51 $\pm$ 0.11 & 0.06 & 0.09 & 0.48 $\pm$ 0.07	\\
SN2005eq  & 0.67 $\pm$ 0.15 & 0.10 &  0.11 & 0.66 $\pm$ 0.15 & 0.10 &  0.11 & 0.67 $\pm$ 0.11\\
SN2005hc  & 0.69 $\pm$ 0.19 & 0.15 & 0.12 & 0.68 $\pm$ 0.19 & 0.15 & 0.12 & 0.71 $\pm$ 0.16 \\
SN2005iq  & 0.48 $\pm$ 0.13 & 0.09 & 0.09 & 0.54 $\pm$ 0.14 & 0.11 & 0.09 & 0.51 $\pm$ 0.10 \\
SN2005ki & 0.45 $\pm$ 0.14 & 0.11 & 0.09 & 0.51 $\pm$ 0.17 & 0.14 & 0.09 & 0.49 $\pm$ 0.14  \\
SN2006bh & 0.37 $\pm$ 0.10 & 0.06 & 0.08 & 0.43 $\pm$ 0.11 & 0.08 & 0.07 & 0.40 $\pm$ 0.07 \\
SN2007bd & 0.55 $\pm$ 0.13 & 0.06 & 0.11 & 0.61 $\pm$ 0.12 & 0.07 & 0.10 & 0.59 $\pm$ 0.10 \\
SN2007on & 0.23 $\pm$ 0.06 & 0.03 & 0.05 & 0.30 $\pm$ 0.07 & 0.05 & 0.05 & 0.28 $\pm$ 0.05 \\
SN2008R & 0.20 $\pm$ 0.06 & 0.04 & 0.05 & 0.27 $\pm$ 0.07 & 0.05 & 0.05 & 0.25 $\pm$ 0.06  \\
SN2008bc  & 0.60 $\pm$ 0.14 & 0.09 & 0.11 & 0.62 $\pm$  0.15 & 0.10 & 0.11 & 0.63 $\pm$ 0.11 \\
SN2008gp & 0.62 $\pm$ 0.13 & 0.06  & 0.11 & 0.65 $\pm$ 0.11 & 0.07 & 0.09 & 0.64 $\pm$ 0.09\\
SN2008hv  & 0.48 $\pm$ 0.11 & 0.07 & 0.09 & 0.54 $\pm$ 0.12 & 0.08 & 0.09 & 0.52 $\pm$ 0.09 \\
SN2008ia  & 0.50 $\pm$ 0.12 & 0.06 & 0.10 & 0.57 $\pm$ 0.11 & 0.07 & 0.09 & 0.55 $\pm$ 0.09  \\
SN2011fe  & 0.50 $\pm$ 0.12 & 0.07 &  0.10 & 0.55 $\pm$ 0.12 & 0.08 & 0.09 & 0.52 $\pm$ 0.10 \\

\hline
\end{tabular}
\label{tab:mni}
\end{center}
\end{table*}

%% file: tab/14j_meth.tex
\begin{table*}
\begin{minipage}{\textwidth}
\begin{center}
\caption{Comparison of different methods to estimate $M_{^{56}Ni}$ for SN~2014J. All measurements assume a distance modulus of $27.64 \pm 0.10$.}
\begin{tabular}{lccl}
\hline
 $M_{Ni}$ (inferred) & $\sigma$ & Method & Reference\\
\hline
0.62	& 0.13 & $\gamma$ ray lines & \citet{Churazov2014} \\
0.56 & 0.10 & $\gamma$ ray lines & \citet{Diehl2015} 
\\
0.37	& $\ldots$ & Bolometric light curve $A_V$=1.7 mag &  \citet{Churazov2014, Margutti2014} \\
0.77	& $\ldots$ & Bolometric light curve $A_V$=2.5 mag & \citet{Churazov2014, Goobar2014a} \\
0.64	& 0.13 & NIR second maximum & this work (combined fit) \\

0.60    & 0.10 & NIR second maximum $+$ measured rise & this work		\\  

\hline
\end{tabular}
\end{center}
\end{minipage}
\label{tab:meth}
\end{table*}

%% file: red_tab.tex
\begin{table*}
\begin{minipage}{\textwidth}
\begin{center}
\caption{\Mni estimates for  objects with high values of $E(B-V)_{host}$. Comparison with independent estimates from the literature are given where available.}
\begin{tabular}{lccccc}
\\
\hline
SN & $t_2$ & \Mni (inferred)  &  \Mni (Lit. Val.) & Percent Difference & Reference \footnote{The references for the \Mni measurements are RL92: \citet{RL1992}, S06a: \citet{Stritzinger2006a}, S06b: \citet{Stritzinger2006b}, St05: \citet{Stehle2005}, ER06: \citet{ER06}, L09: \citet{Leloudas2009}, W08: \citet{Wang2008}, S10: \citet{Scalzo2010}}\\
 & (d) & (\s) & (\s)  &\\

\hline
SN 1986G	& 16.4 $\pm$ 1.4	&	0.33 $\pm$ 0.08 &0.38 $\pm$ 0.03 & 15.15 &RL92  \\
SN 1998bu & 29.9 $\pm$  0.4 &  0.58 $\pm$ 0.12 &  0.57 & 1.7 & S06b\\
SN 1999ac & 27.0 $\pm$ 2.0 &  0.53 $\pm$ 0.12 &  0.67 $\pm$ 0.29 & 26.4 & S06a\\
SN 2001el& 31.2 $\pm$  0.7 &  0.62 $\pm$ 0.12 &  0.40 $\pm$ 0.38 & 33.8 & S06a\\
SN 2002bo & 28.9 $\pm$ 0.7 &  0.56 $\pm$ 0.12 &  0.52 & 7.1 & St05\\
SN 2003cg & 30.2 $\pm$ 1.5 &  0.59 $\pm$ 0.13 &  0.53 & 10.1 & ER06\\
SN 2003hv & 22.3 $\pm$ 0.1 &  0.43 $\pm$ 0.11 &  0.40 $\pm$ 0.11 & 6.9 & L09\\
SN 2006X	& 28.2  $\pm$ 0.5	& 0.57 $\pm$ 0.11 & 0.50 $\pm$ 0.05 & 12.2 & W08 \\
SN 2007if   &   32.3  $\pm$ 0.8 & 0.65 $\pm$ 0.16 & 1.6 $\pm$ 0.1 & 158.3 & S10\\  
\hline

\end{tabular}

\label{tab:red}

\end{center}

\end{minipage}
\end{table*}

%% file: nitable6.tex
\begin{table}
\begin{minipage}{70mm}
\begin{center}
\caption{\Mni\ and \lm measurements for the complete sample of objects with $t_2$ measurements}
\label{tab:yj}
\begin{tabular}{lcccc}
\hline
SN & $M_{^{56}Ni}$ 	& $\sigma$ & $L_{max}$ & $\sigma$ \\
 & \multicolumn{2}{c}{(\s)} & \multicolumn{2}{c}{($10^{43}$ erg/s)}\\
\hline
1980N & 0.42 & 0.10 & 0.84 & 0.21 \\ 
1981B & 0.63 & 0.13 & 1.26 & 0.21\\ 
1986G & 0.33 & 0.07 & 0.66 & 0.18 \\
1998bu & 0.58 & 0.12 & 1.16 & 0.20\\
1999ac & 0.53 & 0.12 & 1.05 & 0.20\\
1999ee & 0.68 & 0.15 & 1.36 & 0.21  \\
2000E & 0.62 & 0.14 & 1.24 & 0.22 \\
2000bh & 0.65 & 0.14 & 1.30 & 0.22 \\
2001bt & 0.55 & 0.12 & 1.10 & 0.20\\
2001cn & 0.58 & 0.13 & 1.19 & 0.20\\
2001cz & 0.67 & 0.14 & 1.33 & 0.22\\
2001el & 0.61 & 0.13 & 1.22 & 0.21\\
2002bo & 0.56 & 0.11 & 1.12 & 0.21\\
2003cg & 0.64 & 0.13 & 1.19 & 0.22\\
2003hv & 0.43 & 0.10 & 0.84 & 0.17 \\
2004ey & 0.57 & 0.14 & 1.14 & 0.20	\\ 
2004gs & 0.43 & 0.11 & 0.85 & 0.18	\\ 
2004gu & 0.71 & 0.17 & 1.42 & 0.23	\\ 
2005A & 0.56 & 0.13 & 1.12 & 0.18	\\ 
2005al & 0.49 & 0.13 & 0.97 & 0.21	\\ 
2005na & 0.64 & 0.15 & 1.28 & 0.22	\\ 
2006D & 0.49 & 0.13 & 0.98 & 0.19	\\ 
2006X & 0.57 & 0.11 & 1.13 & 0.19	\\ 
2006ax & 0.62 & 0.15 & 1.24 & 0.21	\\ 
2006et & 0.64 & 0.16 & 1.27 & 0.22	\\ 
2006gt & 0.39 & 0.09 & 0.77 & 0.18 \\
2006hb & 0.41 & 0.11 & 0.81 & 0.19	\\ 
2006kf & 0.47 & 0.12 & 0.94 & 0.19	\\ 
2007S & 0.71 & 0.16 & 1.41 & 0.22	\\ 
2007af & 0.57 & 0.14 & 1.16 & 0.20	\\ 
2007as & 0.47 & 0.14 & 0.94 & 0.25	\\ 
2007bc   & 0.55 &  0.14 & 1.09 & 0.20 \\
2007bm & 0.54 & 0.13 & 1.08 & 0.20	\\ 
2007ca   & 0.66 &  0.16 & 1.29 & 0.22\\
2007if  & 0.65 & 0.16 & 1.30 & 0.22\\
2007jg   & 0.53 & 0.14 & 1.06 & 0.20 \\
2007le & 0.61 & 0.15 & 1.21 & 0.20	\\ 
2007nq & 0.46 & 0.13 & 0.92 & 0.20	\\ 
2008C & 0.63 & 0.16 & 1.26 & 0.23
	\\ 
2008fp & 0.62 & 0.13 & 1.24 & 0.21	\\ 
2014J & 0.64 & 0.13 & 1.28 & 0.22 \\ 
\hline
\end{tabular}
\end{center}
\label{tab:full}
\end{minipage}
\end{table}

%% file: dis.tex
Using the relation derived from the low-reddening sample we extrapolate an $L_{max}$ value for 58 SNe~Ia objects having a measured $t_2$. The estimate of $t_2$, along with this relation, provides a method to deduce the bolometric peak luminosity, independent of a reddening estimate, distance measurement (relative to the calibration of our low-absorption sample) and without requiring multi-band photometry. We hence have established a reddening-free luminosity function of SNe~Ia at peak (Fig.~\ref{fig:hist}). 

We established an intrinsic luminosity function and \Nif mass distribution for all SNe~Ia with a $t_2$ measurement (Tab.~\ref{tab:full}). 
The distribution of \lm has a standard deviation of 0.2 $\cdot 10^{43}$ erg s$^{-1}$ and \Mni has a standard deviation of 0.11\,M$_\odot$. \citet{Scalzo2014} find a similar distribution of \Mni with a $\sigma$ of 0.16\,M$_\odot$.  
We test our method on SN~2014J, a heavily reddened SN~Ia in the nearby galaxy M82 and find good agreement between the estimates from the $\gamma-$ray observations \citep[][see Table~\ref{tab:meth}]{Churazov2014, Diehl2015}.
Faint, 91bg-like SNe~Ia, which show typically lower luminosities \citep{Filippenko1992,Leibundgut1993}, do not display a second maximum in their NIR light curves and are not in our sample. Therefore, the true dispersion, in peak luminosity and $M_{^{56}Ni}$, for SN~Ia  will likely be larger than what is derived here. \citet{Stritzinger2006a} find a dispersion of a factor of $\sim$ 10, since their sample included peculiar SNe~Ia like SN~1991bg and SN~1991T.

Our reddening-free estimate of the \Mni can be compared to independent \Nif mass estimates, e.g. from the late-time ($\geq$ 200\,d) pseudo-bolometric light curve. It should also be possible to determine the amount of radiation emitted outside the UVOIR region of the spectrum at late phases and a bolometric correction \citep[e.g.][]{Leloudas2009}. There are very few objects for which both NIR data to measure $t_2$ and nebular phase pseudo-bolometric observations are present, making a quantitative comparison for a sample of objects extremely difficult. Thus, we strongly encourage more late-time observations of SN~Ia.

The observed \lm and \Mni distributions directly connect to the physical origin of the diversity amongst SNe~Ia.
A possible explanation is the difference in the explosion mechanism. Pure detonations of $M_{ch}$ WDs \citep{Arnett1969} were seen to be unfeasible since they burn the entire star to iron group elements and do not produce the intermediate mass elements (IMEs) observed in SN~Ia spectra. Pure deflagrations \citep[e.g.][]{Travaglio2004} can reproduce observed properties of SNe with \Mni $\leq$ 0.4\,M$_\odot$. Deflagration models however, cannot account for SNe with higher \Mni and hence, cannot explain the entire distribution in Figure~\ref{fig:hist}.

Delayed-detonation models \citep[e.g.][]{Khokhlov1991,Woosley1990} are more successful in producing higher M$_{^{56}Ni}$. In this explosion model a subsonic deflagration expands the white dwarf to create low densities for IMEs to be produced in a supersonic detonation phase which is triggered at a deflagration-to-detonation transition density ($\rho_{tr}$).

Recent 1D studies by \citet{Blondin2013} confront a suite of Chandrasekhar mass ($M_{Ch}$) delayed detonation models with observations for SNe with a range of peak luminosities. They find a very good agreement of their models with photometric and spectroscopic observations at maximum. The range of \Mni produced by their models corresponds well with the observations in Figure~\ref{fig:hist}, making these models a strong candidate to explain the observed diversity. 
 
\mc explosion models can possibly account for the observed distribution in M$_{^{56}Ni}$.
Recent studies \citep[e.g.][]{VK2010} on the other hand posit sub-Chandrasekhar mass
explosions as a progenitor scenario for SNe~Ia \citep[for e.g., see][]{WW94}. This scenario is attractive since it can account for the progenitor statistics from population synthesis \citep[see][]{Livio2000, Ruiter2013}. Moreover, studies like \citet{Stritzinger2006a} and \citet{Scalzo2014} find a significant fraction of SNe~Ia to have $M_{ej} <$ 1.4\,M$_\odot$, providing observational evidence for the sub-\mc progenitor scenario. 
We compare the luminosity function in Figure~\ref{fig:hist} to the one obtained by \citet{Ruiter2013}, using their violent merger models. They present a relation between primary white dwarf mass ($M_{WD}$) and peak brightness for a grid of sub-$M_{Ch}$ models. For objects in the lowest two bins of our luminosity distribution, the $M_{WD}$ corresponds to 1 to 1.1\,$_\odot$. For the highest luminosity objects, the models indicate an $M_{WD}$ of 1.28\,M$_\odot$. Thus, the luminosity function corresponds to a range of sub-Chandrasekhar $M_{WD}$, which provides further evidence for the plausibility of sub-$M_{Ch}$ explosions as a progenitor scenario. The \Nif mass distribution (Fig.~\ref{fig:hist}) is comparable to the yields from the models of \citet{Sim2010}.   Our \lm and \Mni distributions do not allow us to distinguish which explosion mechanism is responsible for the observed variety.

We note that our sample includes one peculiar,  super-\mc event, SN~2007if \citep{Scalzo2010}, with an estimated $M_{^{56}Ni} = 0.65 \pm$ 0.16 \s using our technique. This is significantly lower than the value estimated in \citet{Scalzo2010} of 1.6 $\pm$ 0.1\,M$_\odot$. The $t_2$ estimate for this object is not exceptionally high, indicating a substantial but not exceptional amount of \Nif (similar to 91T-like SNe). One of the possible reasons for this discrepancy could be that the peak luminosity is not just a product of \Nif decay. This idea has been entertained in theoretical models for these super-\mc  SN~Ia. The models  advocate a scenario of ejecta interaction with circumstellar material \citep[CSM; see][]{Hachinger2012, Dado2015}. There is also an indication of a shell interaction in this supernova \citep{Scalzo2010} and if this interaction results in increased peak luminosity then the \Nif mass through Arnett's rule would be overestimated. It could well be that additional energy is emitted in these super-\mc objects. 
A significant, but not extreme, amount of \Nif produced in the explosion along with interaction with the CSM could then explain the observed properties, e.g. lower ejecta velocities ($\sim$ 9000 km s$^{-1}$) and high peak luminosity. In \citet{Hachinger2012}, the lower limit on \Mni is $\sim$ 0.6 \s which agrees well with our estimate. 

The literature for such super-\mc objects with NIR light curves is still limited. Using the data in \citet{T11} for SN~2009dc, we obtain a $t_2$(J) of 31.7 $\pm$ 6.2 \,d which corresponds to an \Mni of $0.65 \pm 0.18\,M_\odot$.
\citet{T13} also argue for less extreme \Mni based on late phase photometry and spectroscopy, although they prefer a comparatively higher \Mni ($\sim$1 $M_\odot$) than our inferred value. One possible reason could be that the high ejecta densities lead to an earlier onset of the recombination wave than expected for normal Ia's and hence an earlier $t_2$ than is expected for a given \Nif mass. This would lead to an inference of lower \Mni from $t_2$ for super-\mc SNe. 

If we assume that the inferred \Nif mass from $t_2$ indicate the core \Nif for all SNe\,Ia the peak luminosity of super--\mc SNe\,Ia would be boosted by an additional energy source, like shell interaction within the explosion. A good indicator could be the late bolometric decline phase and luminosity. This comparison would be much closer to the \Nif determitation of the second peak than the peak bolometric luminosity.

Larger samples of well-observed SNe \citep[e.g.][]{Friedman2015} will help in improving the statistics of such a study. Future investigations with a detailed comparison between observations and a suite of sub-\mc detonation models will help 
shed more light on the nature of the progenitor scenario and explosion mechanism of SN~Ia. Moreover, 
future theoretical studies of peculiar, super-\mc SNe  will help in deciphering the nature of these extreme explosions.